# Thermoelectric signature of individual skyrmions

*Alexander Fernández Scarioni [1]\*, Craig Barton [2], Héctor Corte-León [2], Sibylle Sievers [1], Xiukun Hu [1], Fernando Ajejas [3], William Legrand [3], Nicolas Reyren [3], Vincent Cros [3], Olga Kazakova [2], and Hans W. Schumacher [1]*

1)      Physikalisch-Technische Bundesanstalt, 38116 Braunschweig, Germany

2)      National Physical Laboratory, Teddington TW110LW, United Kingdom

3)      Unité Mixte de Physique, CNRS, Thales, Université Paris-Saclay, 91767 Palaiseau, France

**Abstract**

We experimentally study the thermoelectrical signature of individual skyrmions in chiral Pt/Co/Ru multilayers. Using a combination of controlled nucleation, single skyrmion annihilation, and magnetic field dependent measurements the thermoelectric signature of individual skyrmions is characterized. The observed signature is explained by the anomalous Nernst effect of the skyrmion's spin structure. Possible topological contributions to the observed thermoelectrical signature are discussed. Such thermoelectrical characterization allows for non-invasive detection and counting of skyrmions and enables fundamental studies of topological thermoelectric effects on the nano scale

Skyrmions are nano-scale topologically non-trivial spin structures which are inherently robust due to their particular topology and can be driven efficiently by electrical currents [1-3], with potential applications in logic and storage devices and neuromorphic computing [4]. Their electrical characterization and manipulation have been investigated intensely over recent years [5-14] leading to observations such as the skyrmion Hall effect [15,16] and the topological Hall effect [14,17]. However, only very few studies have addressed or exploited their thermoelectrical properties [18-21]. While these studies mainly address the topological properties of bulk materials, in thin films the thermoelectrical signal has recently been used as a tool to image skyrmions [21]. In conventional spintronic materials thermoelectric studies have led to important discoveries such as the tunneling magneto-Seebeck effect [22] or spin heat accumulation [23] and have enabled nano-scale detection of domain wall motion [24]. For skyrmions such thermoelectric studies could shine light on topological contributions to the Nernst effect [25] and might provide new tools for skyrmion detection and manipulation.

In this Letter, we experimentally study the thermoelectrical signature of individual skyrmions in a Pt/Co/Ru multilayer micro stripe and attribute it unambiguously to the anomalous Nernst effect (ANE) originating from the spin structure of the skyrmion without significant topological contributions. Our experiments show that beyond this fundamental insight into the thermoelectrical properties of individual nano-scale topological spin structures our findings enable non-invasive characterization, detection and counting of skyrmions in magnetic microdevices.

We use micron-wide stripes made of Pt/Co/Ru multilayers with perpendicular magnetic anisotropy (PMA) and strong interfacial Dzyaloshinskii-Moriya interaction (DMI) as shown in Fig. 1(a). The multilayer stack is comprised of asymmetric trilayers based on Ta(5)/Pt(8)/[Co(1)/Ru(1.4)/Pt(0.6)]x10/Pt(2.4) (numbers are thickness in nm) deposited on a thermally oxidized Si substrate (methods in [26]). The microstripes were patterned by electron beam lithography and argon etching. Sputter deposited contacts and a 200 nm wide microheater of 5 nm Ta and 95 nm Pt were patterned by lift-off. The microheater is used to induce a transverse in-plane thermal gradient $\nabla T_x$ in the microstripe. $\nabla T_x$ is perpendicular to the PMA magnetization **M** resulting in an ANE voltage $V_{ANE}$ in the longitudinal direction between the two contacts which can be expressed by:

$$V_{ANE} = -N_{ANE}\mu_0 l \overline{\nabla T_x} M_z \quad (1).$$

Here $N_{ANE}$ is the ANE coefficient per magnetic moment, $\mu_0$ is the vacuum permeability, $l$ is the length of the sensed region, $\overline{\nabla T_x}$ is the averaged x-component of the thermal gradient across the material. and $M_z$ is the averaged z-component of the magnetization in the sensing region between the measurement contacts [Fig. 1(a)]. In our study the saturation condition is denoted by $M_z = M_s$, the saturation magnetization.

The ANE in the lithographically patterned devices was first characterized by magnetization reversal hysteresis loops. In the ANE voltage measurements, $V_{ANE}$ was detected at the second harmonic using a lock-in technique [27]. Fig. 1(b) shows a hysteresis loop of $V_{ANE}$ as function of the perpendicularly applied out-of-plane field $\mu_0 H_z$. The z component of the averaged magnetization, $M_z$, shows the expected reversal behavior of PMA samples with maze domain as observed in the non-patterned thin film by Magnetic Force Microscopy (MFM) at zero applied magnetic field [inset of Fig. 1(h)]. With increasing applied field, regions of reversed magnetization expand until full saturation is reached. as can be seen in MFM micrographs of the microstripe in different applied fields [Fig. 1(c)-1(e)]. Additionally, individual skyrmions (white arrow in Figure 1e), occur, demonstrating field stabilization of skyrmions during the reversal loops.

To extract the temperature distribution and $\overline{\nabla T_x}$, finite element heat flux modelling of the device under the experimental conditions was performed [27]. The resulting temperature distribution for a dc microheater current $I_{heater}$ of 4.3 mA is shown in Fig. 1(f). The average thermal gradients along the x,y,z axes in the device are found to be $\overline{\nabla T_x}$ = 4.2 K/µm, $\overline{\nabla T_y}$ = 2.9 mK/µm and $\overline{\nabla T_z}$ = 0.2 mK/µm, respectively. Given the order of magnitude difference between $\overline{\nabla T_x}$, and $\overline{\nabla T_y}$ as well as $\overline{\nabla T_z}$, we neglect the latter two contributions to the ANE in further analysis. In Fig. 1(g), we show the simulated surface temperature profile across the microstripe along the dashed line in Fig. 1(f). Inside the microstripe (inset), the temperature $T_{stripe}$ increases by about 8.6 K. Using the microstripe as a resistive thermometer, the averaged $T_{stripe}$ can be experimentally determined as function of $I_{heater}$ [27]. As shown in Figure 1h (black circles), $T_{stripe}$ increases quadratically with $I_{heater}$ as expected for Joule heating and the experimentally determined temperature increase of 8.6 K for the maximum $I_{heater}$ = 4.3 mA agrees well with the simulations.

Similar ANE hysteresis loops as in Figure 1b were measured for a range of input heat currents $I_{heater}$ = 0.2 - 4.3 mA. In Figure 1h, we show the ANE amplitude $V_{ANE}^{max}$ (red squares) derived from subtracting the measured saturated $V_{ANE}^{sat}$ for positive and negative fields. Again, $V_{ANE}^{max}$ scales quadratically with $I_{heater}$ and thus proportionally with $\overline{\nabla T_x}$ confirming the thermoelectrical origin. Using Eq. (1), the measured $V_{ANE}^{max}$ with the average $\overline{\nabla T_x}$ from the temperature calibration/simulations and $\mu_0 M_s$ = 1.4 T from magnetometry data, a Nernst coefficient of $N_{ANE}$ = 8.1 nV/KT is derived.

To controllably study the thermoelectrical response of few or *individual* skyrmions we use a combination of current induced nucleation [8-13] and magnetic fields to generate skyrmions in the microstripe, and image them by in-situ MFM [27]. In the MFM data dark contrast corresponds to negative out-of-plane magnetization [cp. coordinate system in Fig. 1(a)].

MFM images of the microstripe before and after driving a single rectangular current pulse with nominal duration of 200 ns through the microstripe are displayed in Fig. 2(a)-2(d). Applied pulse current densities j ranged from 3.34 x $10^{11}$ A/m$^2$ to 3.97 x $10^{11}$ A/m$^2$ and the applied magnetic fields $\mu_0 H_z$ = 0.49 - 40.7 mT. Before the pulse, an initial maze domain state was obtained by first saturating the sample at negative field and then sweeping to the desired positive field value. Depending on the relative intensities of j and $\mu_0 H_z$ various end states comprised of maze domains and skyrmions were found that could be classified into the following three categories: pure stripe domains [Fig. 2(a)], coexistence of skyrmions and stripe domains [Fig. 2(b)], or only single skyrmions [Fig. 2(c) and 2(d)], as summarized in the nucleation phase diagram, Fig. 2(e). The dotted line marks the region where single digit numbers of skyrmions are nucleated (blue triangles). Repeated measurements did not reveal preferential nucleation sites but rather stochastic skyrmion distributions in the microdevice indicating a thermally driven nucleation process.

Although the above protocol allows us to generate single digit numbers of skyrmions it does not allow us to controllably change the number of skyrmions. Therefore, we applied a single skyrmion annihilation procedure exploiting the confined stray magnetic field around the MFM tip apex [28-29]. In Fig. 3(a), we show an MFM image after nucleation of five skyrmions in an applied magnetic field of +34.84 mT. For a given external applied magnetic field, i.e. +11.3 mT, the local total field (the sum of the external applied field and the local tip stray field) under the MFM tip is below the skyrmion annihilation field (between 50 mT and 60 mT). This allows us to image the skyrmions without detrimental tip-sample interaction. In contrast, when the external field exceeds +24 mT, the total local field under the MFM tip exceeds the annihilation threshold. The targeted application of external fields thus allows to controllably delete individual nanoscale skyrmions when placing the MFM tip above them. Two examples of this selective annihilation process are shown in the MFM micrograph sequences, Fig. 3(a)-3(c) and Fig. 3(d)-3(f).

Importantly, our experimental setup allows *in-situ* MFM characterization and annihilation of the skyrmions simultaneously with the thermoelectrical characterization. This is highly advantageous as it allows, for the first time, the measurement of the thermoelectrical signature of individual skyrmions, analogously to previous studies of the thermoelectrical signature of an individual magnetic domain wall in a nanowire [24] or of the anomalous Hall effect detection of skyrmions [11]. The measured ANE signal from the skyrmion configurations shown in Fig. 3(a)-3(c) and Fig. 3(d)-3(f) is plotted in in Fig. 3(g), red dots and blue triangles, respectively. The difference $\Delta V_{ANE}$ of the measured ANE signal to the ANE in saturation ($\Delta V_{ANE} = V_{ANE} - V_{ANE}^{sat}$) is plotted as a function of the number of detected skyrmions in the sensing region. The data point for zero skyrmions corresponds to the saturated signal $V_{ANE}^{sat}$ where, by definition, $\Delta V_{ANE} = 0$. The error bars reflect the standard deviation of the measurement noise of each configuration. $\Delta V_{ANE}$ shows a clear linear dependence on the skyrmion number. By linear regression with fixed zero intercept an average ANE voltage signature per skyrmion of $V_{ANE}^{sky} = 4.6 \pm 0.2$ nV is derived for the given experimental conditions, i.e. $\mu_0 H_z = +11.3$ mT, $\overline{\nabla T_x} = 4.2$ K/μm.

Following Eq. (1), the measured $\Delta V_{ANE}$ is proportional to the net $M_z$ in the area between the two measurement contacts, and hence proportional to the cumulative skyrmion induced change of the magnetization. The observed reduction of $\Delta V_{ANE}$ with the number of skyrmions should then scale with the net reversed magnetization area inside the skyrmions rather than with the skyrmion number. We therefore estimate an effective skyrmion area $A_{sky}$ of each individual skyrmion from MFM measurements by fitting a 2D elliptical Gaussian function with arbitrary in-plane orientation of the main axes. $A_{sky}$ is then calculated as an ellipse using the two sigma values of the Gaussian fit [27]. The lower left inset to Fig. 4(a) shows a zoom-in view of an MFM image of a typical skyrmion with the derived skyrmion area $A_{sky}$ marked by the white line. The upper right inset demonstrates a very good agreement of a profile through one skyrmion extracted from an experimental MFM phase shift signal (black) with the Gaussian fit (red). The main panel in Fig. 4(a) displays $\Delta V_{ANE}$ data collected from three nucleation and erasure sequences as function of the total reversed skyrmion area when n skyrmions are present, $A_{tot,n} = \sum_{i=1}^{n} A_{sky,i}$ summed over all n skyrmions inside the microstripe. Again, a linear behavior is observed. The orange line is the calculated ANE signal considering the average thermal gradient $\overline{\nabla T_x}$, the estimated $N_{ANE}$ and the total reversed skyrmion area $A_{tot,n}$. Apparently, the ANE response can be very well described by the reversed skyrmion area without considering any topological contributions. Note that the thermal simulations of Fig. 1(f) suggest a spatial variation of $\overline{\nabla T_x}$ over the width of the wire which would result in dependence of the ANE response on the skyrmion's *x*-position inside the wire which is not observed experimentally. This indicates that the spatial variation of $\overline{\nabla T_x}$ is overestimated by the simulations [27].

Additionally, we investigate the change of the ANE response of a single skyrmion as function of the skyrmion area as tuned by an out-of-plane magnetic field. Fig. 4(b) shows a sequence of five MFM micrographs of a single skyrmion in different fields. The variation $\Delta V_{ANE}$ as function of the skyrmion area of Fig. 4(b) (blue triangles) is plotted in Fig. 4(c) (blue triangles). The red dots and black squares correspond to two additional, separate measurement sequences on different skyrmions [27]. Again, the orange line represents the ANE signal calculated from the skyrmion area and describes well the single skyrmion data within the experimental measurement uncertainty.

It has been observed that when spin polarized electrons transverse a topological non-trivial spin texture, such as a magnetic skyrmion, they can accumulate a Berry phase [30] which can be viewed analogous to an Aharonov-Bohm phase resulting from a fictious magnetic field antiparallel to the uniform magnetization [31]. Experimentally this accumulated Berry phase can result in a topological Hall effect [17] or a topological Nernst effect [25] as observed in the skyrmion phase of bulk MnGe [18] and MnSi [19] as well as in $Mn_{1.8}PtSn$ thin films [20].

In principle, the topological contribution $V_{ANE}^{top}$ to the measured Nernst signal $V_{ANE}$ should scale with the total topological charge of the skyrmions inside the microwire. For the chiral system under investigation, each skyrmion thus is expected to carry the same topological charge $\xi_{top}$. Hence the topological contribution should scale with the number of skyrmions, n, like $V_{ANE}^{top} = n\,\xi_{top}$ and not with their net area $A_{tot,n}$. Thus, in the presence of a significant $V_{ANE}^{top}$ the total measured thermovoltage signal $V_{tot,n} = V_{ANE}(A_{tot,n}) + n\,\xi_{top}$, should not be proportional to $A_{tot,n}$, However, within the experimental uncertainties such nonlinearity indicating a significant $V_{ANE}^{top}$ is not visible in Fig. 4(a) and 4(c).

For a more in-depth analysis, we extract the incremental $\Delta V_{ANE}$ increases $\delta V_{ANE,n} = \Delta V_{ANE,n} - \Delta V_{ANE,n-1}$ being the change of $\Delta V_{ANE,i}$ when annihilating skyrmion n and thus changing the number of skyrmions (and thus the topological charge of the system) by one. It is plotted against the corresponding incremental area change $\delta A_{tot,n} = -(A_{tot,n} - A_{tot,n-1})$. This allows to separate $V_{ANE}$ and $V_{ANE}^{top}$ contributions since:

$$\delta V_{ANE,n} = -N_{ANE}\mu_0 \overline{\nabla T_x} 2M_s \delta A_{tot,n}/w\ +\ 1\cdot \xi_{top} \qquad (2)$$

and thus $\xi_{top}$ can be extracted as an offset. The width of the microstripe is represented by w. Fig 4(d) shows $\delta V_{ANE,n}(\delta A_{tot,n})$ for all data sets in the study. Again, the predicted ANE signal (orange line) well describes the data within the experimental uncertainty. However, by considering a free linear fit to the data, we obtain a nonzero intercept at zero skyrmion area. The presence of a nonzero intercept would indicate a potential topological contribution $V_{ANE}^{top}$ to the skyrmion Nernst signature independent of $A_{sky}$. However, we emphasize that the derived interception of $1.2 \pm 0.6$ nV is about four times smaller than the ANE contribution per skyrmion ($4.6 \pm 0.2$ nV), but numerically different from zero within $2\sigma$ confidence. Based on this dataset, a topological contribution thus cannot be ruled out. However, considering the statistical uncertainty of our data, further studies are required to unambiguously conclude on the nature of the topological Nernst signature of individual skyrmions. One explanation to the rather low contribution may arise from the fact that for the metallic multilayers studied here, both the electron's mean free path and the spin diffusion length are of the order of few nanometers [14,32] and thus significantly smaller than the average skyrmion diameter. The resulting strong scattering might provide a hypothesis for the absence of a strong topological Nernst contribution in our experimental data.

In conclusion, we have measured and characterized the thermoelectrical signature of individual nanoscale skyrmions in Pt/Co/Ru multilayers enabling non-invasive all-electrical detection and counting of skyrmions in magnetic devices. The thermoelectric signature is well explained by ANE of the reversed magnetization area of the individual skyrmions. Only weak indications of an additional topological Nernst contribution are present in the data. Future experiments using material systems with smaller skyrmion diameters or longer mean free paths of the conduction electrons could allow to unequivocally identify the topological Nernst contribution of individual skyrmions.

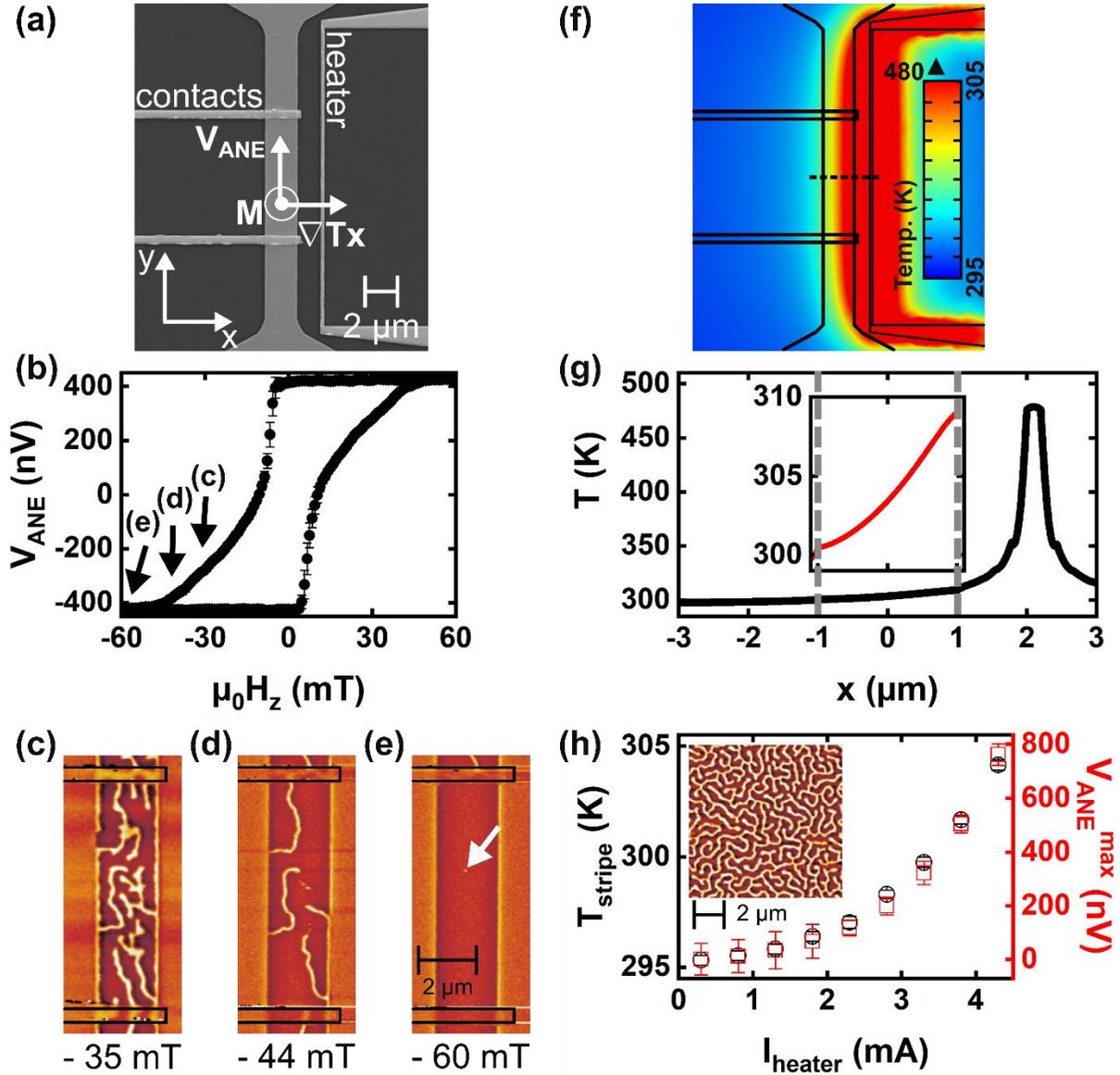

**FIG. 1.** Device characterization and properties. (a) scanning electron micrograph of the device. (b) ANE hysteresis measurement: showing the measured ANE voltage as a function of the applied out-of-plane field. The arrows mark the corresponding domain states imaged by MFM shown in (c–e). (c-e) MFM images of the microstripe for increasingly negative values of the total out-of-plane magnetic field. Here the stated field is the sum of the applied field and the tip stray field at the sample surface. (f) Simulated temperature distribution for $I_{heater}$ = 4.3 mA. (g) Simulated temperature profile along the dashed line in (f). (h) measured temperature rise (black circles) as a function of applied

current and ANE amplitude (red squares) in the microstripe vs. $I_{heater}$. The inset shows an MFM image of the film before patterning in zero field.

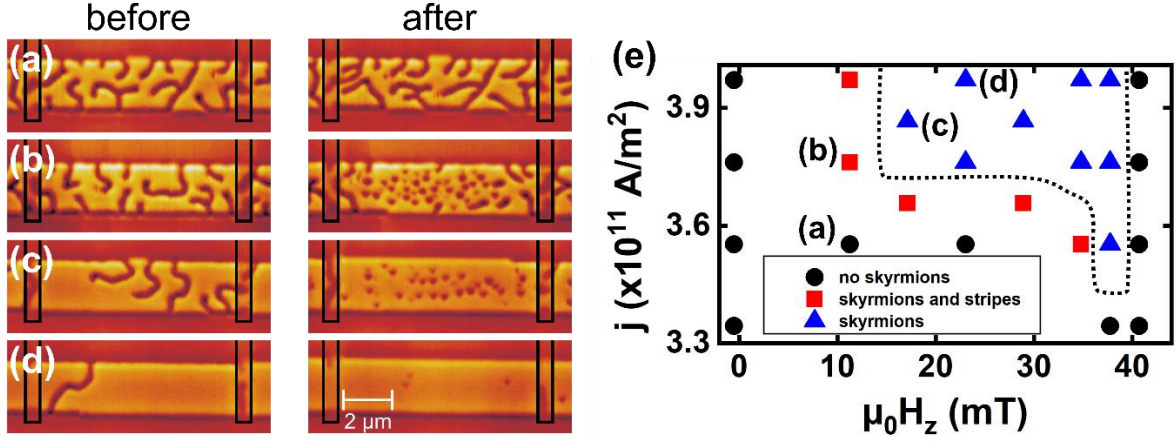

**FIG. 2.** Skyrmion nucleation in the microstripe. (a)-(d) MFM measurements of the microstripe before and after application of a 200-ns-long current pulse. (a) current density j = 3.55 x 10$^{11}$ A/m$^2$, magnetic field $\mu_0 H_z$ = 11.28 mT only perturbation of stripe domains. (b) j = 3.76 x 10$^{11}$ A/m$^2$, $\mu_0 H_z$ = 11.28 mT coexistence skyrmions and stripe domains. (c) j = 3.86 x 10$^{11}$ A/m$^2$, $\mu_0 H_z$ = 17.16 mT only skyrmions (high density). (d) j = 3.97 x 10$^{11}$ A/m$^2$, $\mu_0 H_z$ = 23.05 mT only skyrmions (low density). (e) Nucleation phase diagram for skyrmion generation, the dotted line represents a guide to the eye indicating the parameter range suitable for skyrmion nucleation.

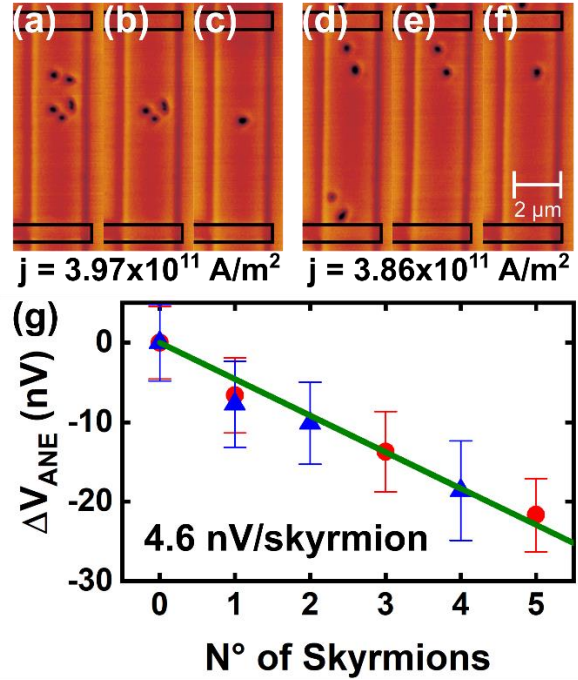

**FIG. 3.** ANE signature of single skyrmions. (a)-(c) and (d)-(f), are two sequences of MFM images demonstrating the probe induced annihilation process of individual skyrmions. (g) $\Delta V_{\mathrm{ANE}}$ as function of the number of skyrmions inside the wire between the ANE contacts corresponding to the skyrmion configurations (a)-(f). The red dots correspond to (a)-(c); the blue triangles correspond to (d)-(f). The green line is a linear fit through the data, yielding an average Nernst voltage of 4.6 ± 0.2 nV per skyrmion.

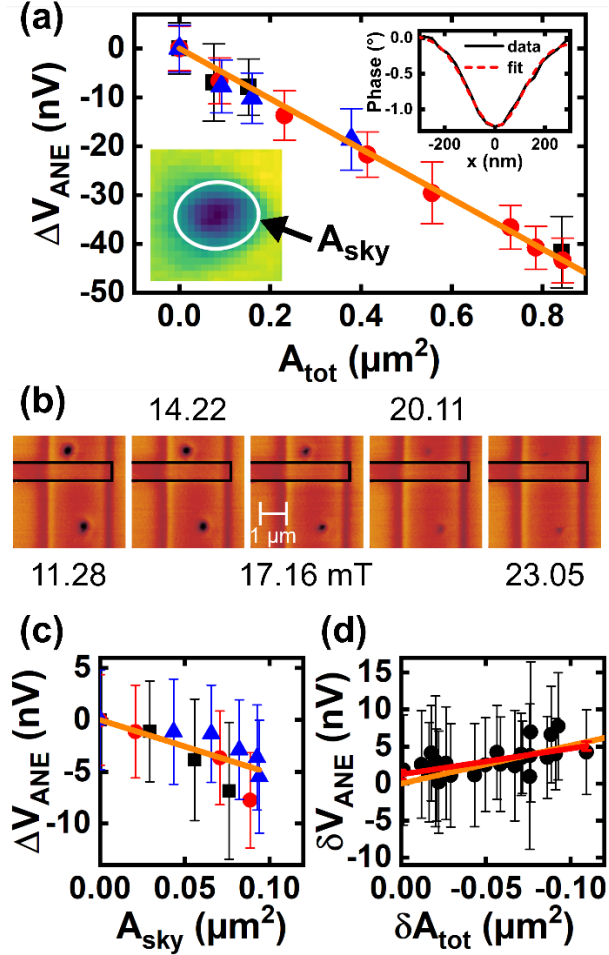

**FIG. 4.** Dependence of the skyrmion ANE signal on skyrmion area. (a) ANE signal as a function of the total effective reversed area. The different symbols correspond to three nucleation/annihilation sequences. The orange line represents the expected ANE voltage calculated for the given reversed magnetization area. Left inset: MFM image of a single skyrmion with the derived $A_{sky}$. Right inset: section through the MFM phase signal (black) of the same skyrmion and Gaussian-fit data (red). (b) MFM showing the field dependence of a single skyrmion. (c) ANE voltage as a function of $A_{sky}$ for single skyrmions from three different nucleation sequences (red dots, blue triangles and black squares) and the ANE signal as calculated from the reversed magnetization area (orange line). (d) Variation in ANE voltage due to the change of the skyrmion number by one. $\delta V_{ANE}$ is plotted against the difference in reversed area. The orange line is the calculated ANE contribution of the reversed area $\delta A_{tot}$. Red line is a linear fit of the data allowing for a non-zero intercept.

**Acknowledgments**

This work was (partially) supported in part by the European Metrology Research Programme (EMRP) and EMRP participating countries under the EMPIR Project No. 17FUN08-TOPS Metrology for topological spin structures.

Financial support from the Agence Nationale de la Recherche, France, under grant agreement No. ANR-17-CE24-0025 (TOPSKY), the Horizon2020 Framework Programme of the European Commission, the DARPA TEE program, through grant MIPR # HR0011831554 under FET-Proactive grant agreement No, 824123 (SKYTOP) is acknowledged.

**References**

[1] Finocchio, G., Büttner, F., Tomasello, R., Carpentieri, M., & Kläui, M. Magnetic skyrmions: from fundamental to applications. Journal of Physics D: Applied Physics 49, 423001 (2016).

[2] Fert, A., Reyren, N., & Cros, V. Magnetic skyrmions: advances in physics and potential applications. Nature Reviews Materials 2, 17031 (2017).

[3] Everschor-Sitte, K., Masell, J., Reeve, R. M., Kläui, M. Perpestive: Magnetic skyrmions- Overview of recent progress in an active research field. Journal of Applied Physics 124,240901 (2018).

[4] Zázvorka, J. et al. Thermal skyrmion diffusion used in a reshuffler device. Nature Nanotechnology 14, 658-661 (2019).

[5] Jiang, W. et al. Blowing magnetic skyrmion bubbles. Science 349, 283-286 (2015).

[6] Woo, S. et al. Observation of room-temperature magnetic skyrmions and their current-driven dynamics in ultrathin metallic ferromagnets. Nature Materials 15, 501-506 (2016).

[7] Moreau-Luchaire, C. et al. Additive interfacial chiral interaction in multilayers for stabilization of small individual skyrmions at room temperature. Nature Nanotechnology 11, 444-448 (2016).

[8] Hrabec, H. et al. Current-induced skyrmion generation and dynamics in symmetric bilayers. Nature Communications 8, 15765 (2017).

[9] Legrand, W. et al. Room-Temperature Current-Induced Generation and Motion of sub-100 nm Skyrmions. Nano Lett. 17, 2703-2712 (2017).

[10] Büttner, F. et al. Field-free deterministic ultrafast creation of magnetic skyrmions by spin-orbit torques. Nature Nanotechnology 12,1040-1044 (2017).

[11] Maccariello, D. et al. Electrical detection of single magnetic skyrmions in metallic multilayers at room temperature. Nature Nanotechnology 13, 233-237 (2018).

[12] Zeissler, K. et al. Discrete Hall resistivity contribution from Néel skyrmions in multilayer nanodiscs. Nature Nanotechnology 13, 1161-1166 (2018).

[13] Lemesh, I. et al. Current-Induced Skyrmion Generation through Morphological Thermal Transitions in Chiral Ferromagnetic Heterostructures. Adv. Mater. 30, 1805461 (2018).

[14] Raju, M. et al. The evolution of skyrmions in Ir/Fe/Co/Pt multilayers and their topological Hall signature. Nature Communications 10, 696 (2019).

[15] Jiang, W. et al. Direct observation of the skyrmion Hall effect. Nature Physics 13, 162-169 (2017)

[16] Litzius, K. et al. Skyrmion Hall effect revealed by direct time-resolved X-ray microscopy. Nature Physics 13, 170-175 (2017)

[17] Neubauer, A. et al. Topological Hall Effect in the A Phase of MnSi. Phys. Rev. Lett. 102, 186602 (2009).

[18] Shiomi, Y., Kanazawa, N., Shibata, K., Onose, Y., & Tokura, Y. Topological Nernst effect in a three-dimensional skyrmion-lattice phase. Phys. Rev. B 88, 064409 (2013).

[19] Hirokane, Y., Tomioka, Y., Imai, Y., Maeda, A., & Onose, Y. Longitudinal and transverse thermoelectric transport in MnSi. Phys. Rev. B 93, 014436 (2016).

[20] Schlitz, R. et al. All Electrical Access to Topological Transport Features in Mn1.8PtSn Films. Nano Lett. 19, 2366-2370 (2019).

[21] Iguchi, R. et al. Thermoelectric microscopy of magnetic skyrmions. Scientific Reports 9, 18443 (2019).

[22] Kuschel, T. et al. Tunnel magneto-Seebeck effect J. Phys. D: Appl. Phys. 52, 133001 (2019)

[23] Dejene, F. K., Flipse, J., Bauer, G. E., & Wees, B. J. Spin heat accumulation and spin-dependent temperatures in nanopillar spin valves. Nature Physics 9, 636-639 (2013).

[24] Krzysteczko, P. et al. Nanoscale thermoelectrical detection of magnetic domain wall propagation. Phys. Rev. B 95, 220410 (2017).

[25]. Mizuta, Y. P., & Ishii, F. Large anomalous Nernst effect in a skyrmion crystal. Scientific Reports 6, 28076 (2016).

[26] Legrand, W. et al. Room-temperature stabilization of antiferromagnetic skyrmions in synthetic antiferromagnets. Nature Materials 19, 34–42 (2020).

[27] See Supplemental Material at DOIXXX for details on methods, finite element modeling and the thermal gradient, device thermometry, estimation of the skyrmion area, skyrmion nucleation experiment and skyrmion field dependency and discussion of uncertainty analysis.

[28] Zhang, S. et al. Direct writing of room temperature and zero field skyrmion lattices by a scanning local magnetic field. Appl. Phys. Lett. 112, 132405 (2018).

[29] Casiraghi, A. et al. Individual skyrmion manipulation by local magnetic field gradients. Communications Physics 2, 145 (2019).

[30] Manna, K., Sun, Y., Muechler, L., Kübler, J., & Felser, C. Heusler, Weyl and Berry. Nature Reviews Materials 3, 244-256 (2018).

[31] Binz, B., & Vishwanath, A. Chirality induced anomalous-Hall effect in helical spin crystals. Physica B: Condensed Matter 403, 1336-1340 (2008).

[32] Nguyen, H. Y., Pratt, W. P., & Bass, J. Spin-flipping in Pt and at Co/Pt interfaces. Journal of Magnetism and Magnetic Materials 361, 30-33 (2014).

# Supplemental Material on:

Thermoelectric signature of individual skyrmions

## Authors:

Alexander Fernández Scarioni, Craig Barton, Héctor Corte-León, Sibylle Sievers, Xiukun Hu, Fernando Ajejas, William Legrand, Nicolas Reyren, Vincent Cros, Olga Kazakova, and Hans W. Schumacher

The Supplemental Material contains information on (S1) methods, (S2) finite element modeling and the thermal gradient, (S3) device thermometry, (S4) estimation of the skyrmion area, (S5) skyrmion nucleation experiment and skyrmion field dependency and (S6) discussion of uncertainty analysis.

## S1) Methods

The ANE voltage measurements, the thermal gradient was generated by ac currents with a frequency of 2024 Hz applied to the microheater by a commercial ac current source (Keithley K6221). $V_{ANE}$ was detected at the second harmonic using a lock-in amplifier (Stanford SR-830). Different frequencies were tested, the frequency of 2024 Hz yielding optimum results for in-situ MFM and ANE measurements.

MFM measurements were performed in a NT-MDT Aura SPM in ambient temperature. Magnetic fields were applied by a built-in electromagnet. Commercial low magnetic moment MFM tips (NT-MDT MFM_LM) were used. Before performing the measurements, the magnetic tip was magnetized in the positive direction out-of-plane with respect to the sample (see Fig. 1(a)), so that the dark contrast areas in our measurements correspond to the negative out-of-plane magnetization and the light parts the opposite direction. The MFM images were taken with an estimated oscillating amplitude of 170 nm of the cantilever and a lift height of 50 nm. In all the MFM measurements related to ANE data a current was applied to the microheater while the MFM images were obtained.

## S2) Finite element modeling and the thermal gradient

Thermal modeling of the microheater was performed using finite element heat flux modelling with COMSOL [1] to estimate the temperature rise and thermal gradient within the ANE device. Temperature dependent bulk material parameters, e.g. thermal conductivity κ and heat capacity Cp, were taken from references [2-6], Table S1. The temperature coefficients of the resistance $\alpha = (1/R)(dR/dT)$ of the microdevice and the microheater were calibrated by temperature dependent resistance measurements, S2.

## S3) Device thermometry

The temperature coefficients of resistance have been calibrated for the Pt heater and the microdevice. Both devices were put in a temperature-controlled air bath unit, where temperatures could be stabilized in the range from 288 K to 314 K. The resistances for both the microdevice and the microheater show a linear dependency as expected in this temperature range [Fig. S1(a)]. The resulting coefficients of resistance $\alpha$ were determined as 8.7 x $10^{-4}$ for the microdevice and 1.5 x $10^{-3}$ for the microheater, respectively. The temperature increase in the microheater and microdevice were experimentally determined for different microheater currents, as shown in Fig. S1(b) for an ambient temperature of 295K. The temperature dependence shows the expected quadratic dependency on the microheater current expected for Joule heating. A current of $I_{microheater}$ = 4.3 mA, as used during the ANE measurements, raises the temperature by 8.06 K and 165.1 K for the microdevice and the microheater, respectively.

## S4) Estimation of the skyrmion area

To determine the skyrmion area a composite Gaussian function was applied:

$$G(x,y) = B * exp\left\{-\left[\frac{[(x-x_c)\cos\theta - (y-y_c)\sin\theta]^2}{2\sigma_x^2} + \frac{[(x-x_c)\sin\theta - (y-y_c)\cos\theta]^2}{2\sigma_y^2}\right]\right\}, \quad (S1)$$

$x_c$ and $y_c$ are the center positions of the Gaussian, $\sigma_x$ and $\sigma_y$ are the standard deviations, $B$ is the amplitude of the Gaussian and $\theta$ is an in-plane rotation angle.

The model was verified by comparison to a calculated MFM phase signal from a simulated skyrmion, following the Tip Transfer Function (TTF) approach [7,8]. The stray field of the skyrmion was calculated using the field transfer method after Schendel et al. [9]. This requires calculation of the skyrmion magnetization distribution $m_x(x,y)$, $m_y(x,y)$, and $m_z(x,y)$ from micromagnetic simulations of a typical skyrmion. The micromagnetic simulations were performed according to [10]. The TTF was obtained for a similar tip as used in the experiment using a well characterized Co/Pt reference sample [11].

For the micromagnetic model the following experimental parameters were used: exchange constant A = 10 pJ/m obtained from domain wall profile as imaged by Lorentz TEM, Dzyaloshinskii-Moriya interaction (DMI) constant D = 1.19 mJ/m$^2$ determined with Brillouin light scattering spectroscopy, saturation magnetization $M_S$ = 1100 kA/m and perpendicular anisotropy constant $K_u$ = 1109 kJ/m$^3$ determined by SQUID measurements. In Fig. S2 the simulated MFM signal cross-section along the x direction of the simulated skyrmion is shown. The data points obtained were then fitted by the Gaussian from Eq. (S1), showing good qualitative agreement experimentally obtained data, thus, verifying our approach. The Gaussian fit is compared with the spatial distribution of the $m_z$ component of the skyrmion depicted in Fig. S2. The resulting relation $r_{sky} \approx \sqrt{2\sigma}$, was used to calculate the skyrmion area as $A_{sky} = 2\pi\sigma_x\sigma_y$.

**S5) Skyrmion nucleation experiment and field dependency of single skyrmion**

Figs. S3(a)-S2(c) show three independent sets of MFM data taken following skyrmion nucleation and subsequent annihilation steps. The nucleation parameters were: 3.97 x 10$^{11}$ A/m$^2$ at +34.84 mT, 3.97 x 10$^{11}$ A/m$^2$ at +34.84 mT and 3.86 x 10$^{11}$ A/m$^2$ at +34.84 mT for images (a), (b) and (c) respectively.

In Fig. S3(a), we show an MFM image after nucleation of eleven skyrmions in an applied magnetic field of +34.84 mT. For a given external applied magnetic field, i.e. +11.3 mT, the local total field (applied field + stray field of the probe) under the MFM probe is always below the skyrmion annihilation field (between 50 mT and 60 mT). As such this allows us to image the skyrmions without detrimental probe-sample interaction. By exceeding +24 mT, the total local field under the MFM probe becomes larger than the annihilation field. This allows a precise methodology to annihilate individual skyrmions.

With a single skyrmion in the microdevice, as shown by the final image in the sequences, Figs. S3(a)-S3(c), the magnetic field was increased to investigate the single skyrmion field dependency. In Figs. S4(a)-S5(c) we show the corresponding MFM measurements in the different applied fields.

## S6) Discussion of uncertainty analysis

The Nernst coefficient presented was calculated using an average thermal gradient along the x direction and assuming each skyrmion senses this same thermal gradient. Interestingly, our FEM simulations suggest that the thermal gradient deviates from the constant approximation along the x-direction as plotted in Fig. S5(a). According to the ANE equation, such spatially varying gradient should result in an x-position dependence of the skyrmion's ANE voltage. However, the simple model of a constant gradient inside the sensing region [orange line in Fig. 4(a) in the manuscript] shows a very good agreement with the experimental data.

The upper limits in the uncertainty have been estimated using the FEM and MFM simulations . For each skyrmion an effective local thermal gradient was obtained by averaging over its area using the FEM simulations with a standard error of 10% in position and shape. Additionally, an error is allowed for the skyrmion area. We assume an error in the relationship between the σ and the effective skyrmion radius ($r_{sky} \approx \sqrt{2}\sigma$) of approximately 12% ($r_{sky} \approx \sqrt{(2 \pm 0.25)\sigma}$) as shown Fig. S5. (b). We can now calculate the error in $V_{ANE}$:

$$V_{ANE} = -N_{ANE}\, \mu_0 M_z \overline{\nabla T_{x;local}}\, A_{tot}/w \qquad (S2)$$

where $\overline{\nabla T_{x;local}}$ represents the local average thermal gradient and $w$ = 2 µm the width of the microdevice. Fig. S5(c) shows the comparison between the measured $V_{ANE}$ in Fig. 4(a) (red dots) and the simulated ANE voltage taking in to account the discussed error margins, shown here as the shaded region. The simulated range of responses describes the observed experimental results well as evidenced by the agreement with the superimposed dataset.

**Supplementary information figures and tables:**

| Material | $C_p$ (J/kg K) | $\rho$ (kg/m³) | $\kappa$ (W/m K) | $\alpha$ (1/K) |
|---|---|---|---|---|
| Si | $C_{p;Si}$ (T) [7] | 2329 | $\kappa_{Si}$ (T) [8] | |
| SiO$_2$ | $C_{p;SiO2}$ (T) [9] | 2203 | $\kappa_{SiO2}$ (T) [10] | |
| Pt | $C_{p;Pt}$ (T) [11] | 21450 | | 0.0015 |
| Microdevice | 130 | 8000 | | 8.7e-4 |

**Table S1:** Material parameters used in the FEM simulations.

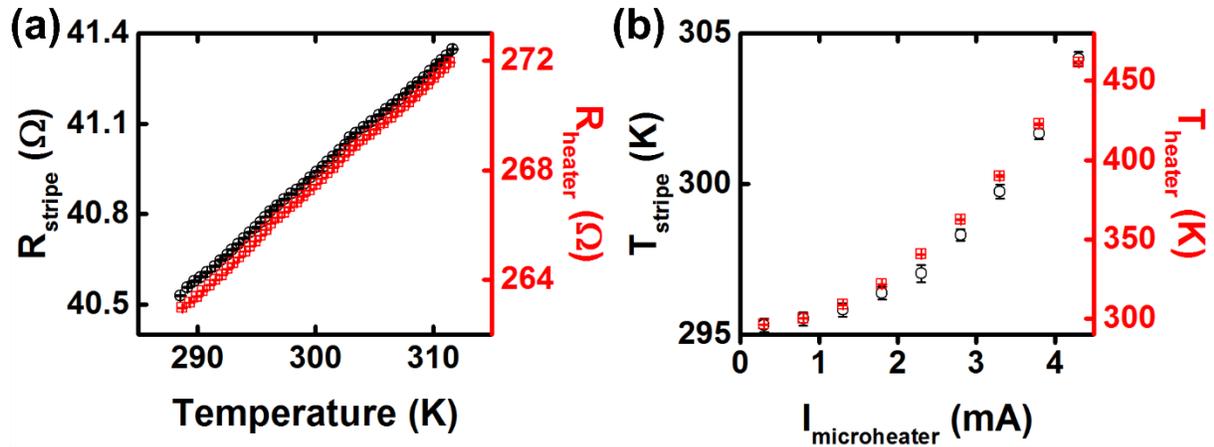

**Figure S1:** Temperature coefficient of resistance calibration. (a) 4-wire resistance measurements of the microdevice (black circles) and the microheater (red squares) as a function of temperature. (b) Temperature of the microdevice (black circles) and microheater (red squares) as a function of I$_{microheater}$.

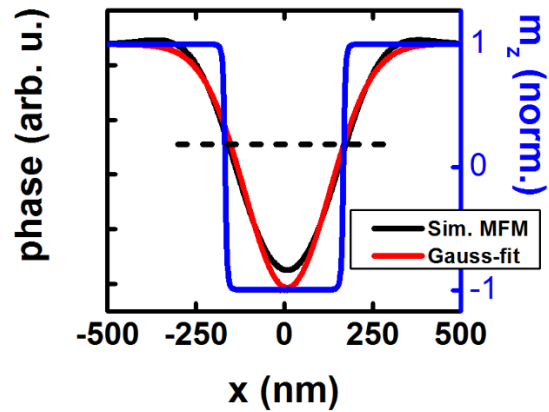

**Figure S2:** Skyrmion magnetization profile comparison with simulated MFM response. x cross-section through the simulated MFM phase data (black line) from a micromagnetic simulated skyrmion, Gaussian-fit data (red line) and the $m_z$ component of the skyrmion used in the simulations (blue line). Dashed black line indicates the cut off criterium used to calculate the skyrmion area representing $\sqrt{2}\sigma$.

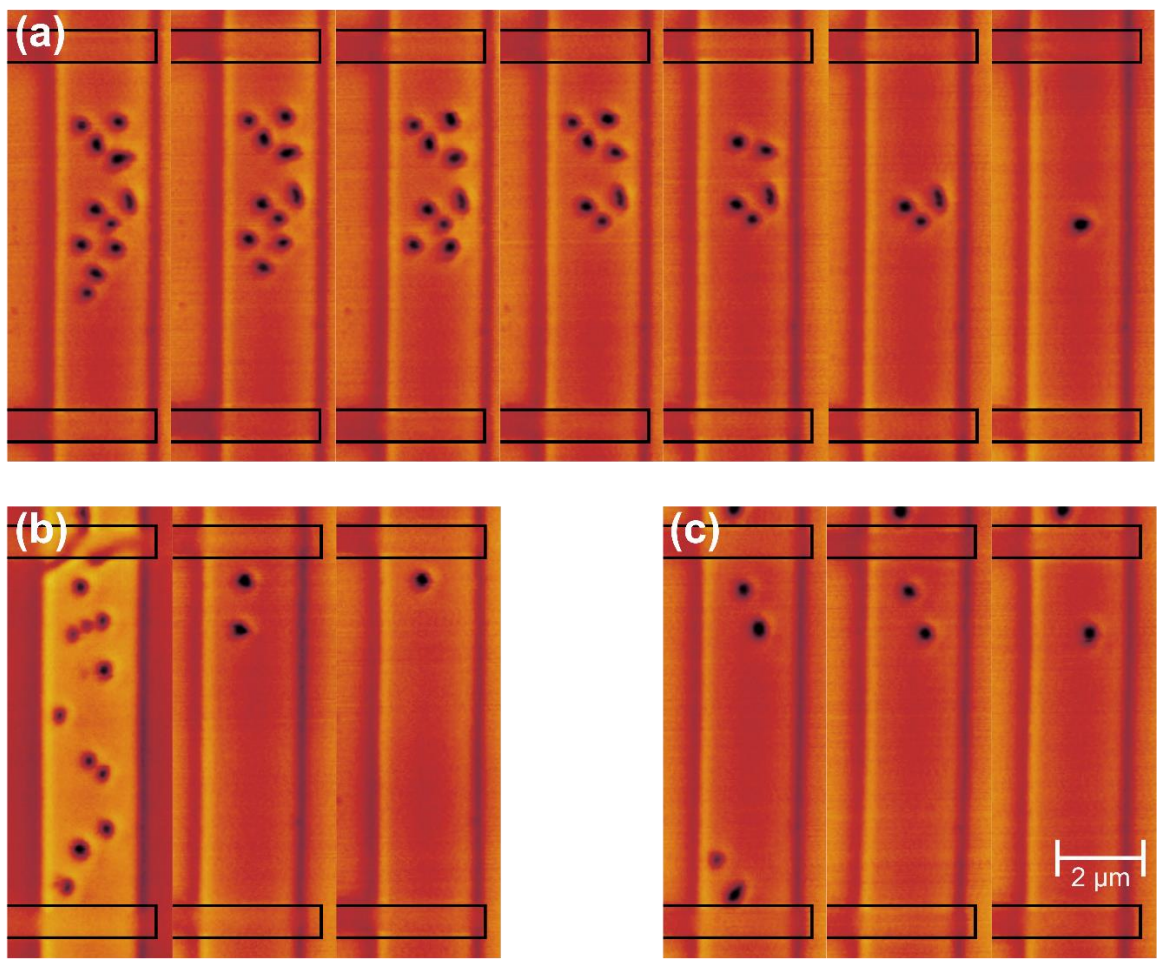

**Figure S3:** Precision annihilation of skyrmions at the single digit level. (a)-(c) MFM measurements of the three independent sequences of skyrmion nucleation (leftmost image) and annihilation by the MFM probe (subsequent images). The applied field is +11.28 mT. All the measurements were performed in the same microdevice.

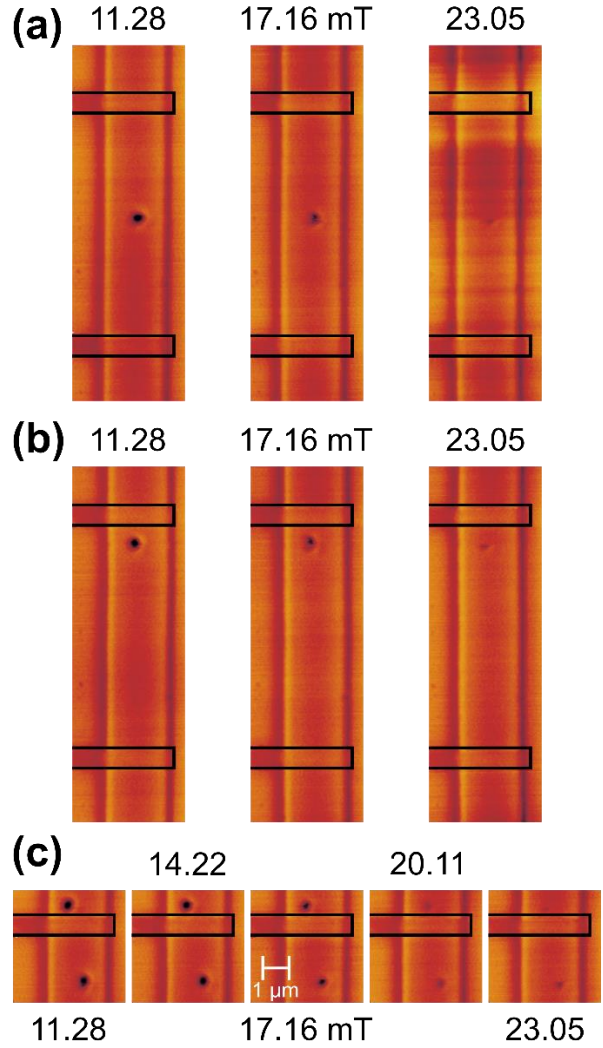

**Figure S4:** Skyrmion size tuning through field dependence measurements. (a)-(c) MFM images of a single skyrmion with increasing the applied field for three independent nucleation sequences.

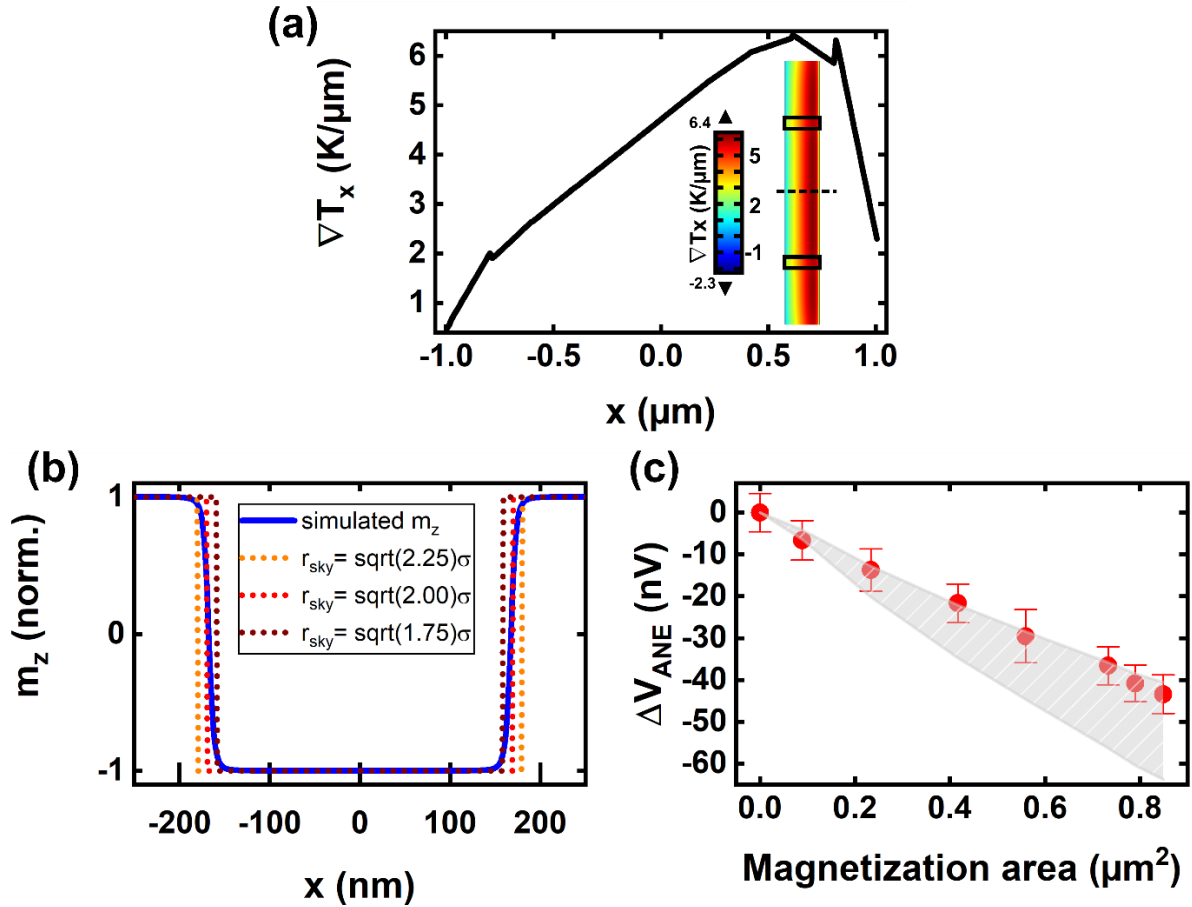

**Figure S5:** Uncertainty analysis in the observed ANE response. (a) cross-section of the temperature gradient along the x direction. The colormap represents the temperature gradient distribution along the x axis of the mircrostripe. (b) the $m_z$ component of the skyrmion used in the simulations (blue line), the different dotted lines represent the radius calculated by $r_{sky} \approx \sqrt{x\sigma}$ where $x$ is the relation that is obtained by comparing the simulated MFM image with the $m_z$ component of the skyrmion. (c) measured ANE voltage as a function of the total effective reversed area (red dots) obtained from the nucleation and annihilation sequence depicted in Fig. S3(a). The gray area depicts the calculated ANE voltage uncertainty.

# References

[1] COMSOL Multiphysics® v. 5.0. www.comsol.com. COMSOL AB, Stockholm, Sweden.

[2] Okhotin, A.S. Pushkarskii, A.S. & Gorbacher, V.V. Thermophysical properties of semiconductors. Atom Publ. House, Moscow, 1972.

[3] Glassbrenner, C. J., & Slack, G. A. Thermal Conductivity of Silicon and Germanium from 3°K to the Melting Point. Phys. Rev. 134, A1058 (1964).

[4] B. S. Hemingway. Quartz: heat capacities from 340 to 1000 K and revised values for the thermodynamic properties. American Mineralogist 72, 273-279 (1987).

[5] Touloukian, Y.S., Powell, R.W., Ho, C.Y., & Klemens, P.G. Thermophysical properties of matter, Vol. 2, IFI/Plenum, New York, p.193 (1970).

[6] Furukawa, G. T., Reilly, M. L., & Gallagher, J. S. Critical Analysis of Heat-Capacity Data and Evaluation of Thermodynamic Properties of Ruthenium, Rhodium, Palladium, Iridium, and Platinum from 0 to 300K. A Survey of the Literature Data on Osmium. Journal of Physical and Chemical Reference Data 3, 163-209 (1974).

[7] Nečas, D. et al. Determination of tip transfer function for quantitative MFM using frequency domain filtering and least squares method. Scientific Reports 9, 3880 (2019).

[8] Hu, X., Dai, G., Sievers, S., Neu, V., & Schumacher, H. W. Uncertainty Propagation and Evaluation of Nano-Scale Stray Field in Quantitative Magnetic Force Microscopy Measurements. 2018 Conference on Precision Electromagnetic Measurements (CPEM 2018), Paris, 2018, pp. 1-2.

[9] Schendel, P. J., Hug, H. J., Stiefel, B., Martin, S., & Güntherodt, H.-J. A method for the calibration of magnetic force microscopy tips. Journal of Applied Physics 88, 435-445 (2000).

[10] Legrand, W. et al. Modeling the Shape of Axisymmetric Skyrmions in Magnetic Multilayers. Phys. Rev. Applied 10, 064042 (2018).

[11] Kazakova, O. et al. Frontiers of magnetic force microscopy. Journal of applied Physics 125, 060901 (2019)